\documentclass[twocolumn, aps, prc, superscriptaddress]{revtex4}
\usepackage{amsmath}
\usepackage{amsfonts}
\usepackage{amssymb}
\usepackage{graphicx}
\usepackage[hidelinks, unicode=true]{hyperref}
\hypersetup{linkcolor=black, citecolor=black, filecolor=black, urlcolor=black}
\usepackage[all]{hypcap}
\usepackage{fourier}

\begin{document}

\title{Proton Cumulants and Correlation Functions in Au + Au Collisions at \texorpdfstring{$\sqrt{s_\text{NN}}$}{sqrt (s (NN)) }=7.7--200 GeV from UrQMD Model}

\author{Shu He}
\affiliation{Key Laboratory of Quark\&Lepton Physics (MOE) and Institute of Particle Physics,\\
Central China Normal University, Wuhan 430079, China}
\author{Xiaofeng Luo}
\email{xfluo@mail.ccnu.edu.cn}
\affiliation{Key Laboratory of Quark\&Lepton Physics (MOE) and Institute of Particle Physics,\\
Central China Normal University, Wuhan 430079, China}
\affiliation{Department of Physics and Astronomy, University of California, Los Angeles, California 90095, USA}

\begin{abstract}
We studied the acceptance dependence of proton cumulants (up to fourth order) and correlation functions in 0--5\% most central Au+Au collisions at $\sqrt{s_\text{NN}}$=7.7, 11.5, 19.6, 27, 39, 62.4 and 200 GeV from UrQMD model.  We found that high order proton cumulants show suppressions at large acceptance. By decomposing the proton cumulants into linear combination of multi-proton correlation functions, we observed the two-proton correlation functions always show negative values due to the effects of baryon number conservations. The three and four-proton correlation functions are close to zero and show negligible acceptance dependence. We further observed that the proton cumulants and correlation functions follow similar trends and show a scaling behavior when plotting the results versus mean number of protons.  The comparisons between experimental data and the UrQMD calculations show that the non-monotonic energy dependence of proton correlation functions measured by STAR experiment cannot be described by the UrQMD model. The UrQMD calculations can provide us baselines for the experimental studies of the proton cumulants and correlation functions. Finally, we propose to measure the rapidity dependence of the reduced proton correlation functions to search for the QCD critical point in heavy-ion collisions.\end{abstract}

\maketitle

\section{Introduction}
Exploring the phase structure of the QCD matter is one of the main goals of the relativistic heavy-ion collision experiment. Lattice QCD calculations show that the transition from hadronic phase to Quark-Gluon Plasma (QGP) phase is a crossover~\cite{crossover} at zero baryon chemical potential ($\mu_\text{B}=0$), while at large $\mu_\text{B}$, QCD based model calculations suggest that the phase transition is first-order~\cite{location,bowman2009critical}. If it is true, there should exist a so-called QCD critical point as an end point of the first order phase boundary~\cite{QCP_Prediction,science,location}. Due to the sign problem in Lattice QCD calculations at finite baryon density, it still has large uncertainties in locating the critical point from the theoretical side~\cite{location,qcp_Rajiv}. 

Fluctuation of conserved quantities, such as net-baryon ($B$), net-charge ($Q$) and net-strangeness ($S$) number, are served as the sensitive observable in heavy-ion collisions to search for the QCD phase transition and critical point~\cite{ejiri2006hadronic, qcp_signal, Neg_Kurtosis, Schaefer:2011ex, Asakawa}. Those have been extensively studied experimentally~\cite{2010_NetP_PRL,netcharge_PRL,STAR_BES_PRL,Luo:2015ewa} and theoretically~\cite{HRG_Karsch, PBM_netpdis, Lattice,2015_JianDeng_fluctuation, chen2016robust, 2014_Bengt_flu, Asakawa_formula, Kitazawa:2012at, BFriman_EPJC, 2015_Swagato_evolution, Mukherjee:2016kyu, 2015_Vovchenko,baseline_PRC, huichao, Jiang:2015cnt, HRG_Nahrgang, kenji_morita, freezeout, Borsanyi:2013hza, Alba:2014eba, Fan:2016ovc}. 
The experimental methods of measuring the conserved quantity fluctuations have been well developed in recent years~\cite{Luo:2017faz}. The STAR experiment has measured the fluctuations of net-proton (proxy for net-baryon), net-charge and net-kaon (proxy for net-strangeness) numbers in Au+Au collisions at $\sqrt{s_\text{NN}}$=7.7, 11.5, 14.5, 19.6, 27, 39, 62.4 and 200 GeV. The data is collected in the first phase of the beam energy scan (BES-I, 2010--2014) program at the Relativistic Heavy-ion Collider (RHIC).

In this work, we performed systematic acceptance dependence studies on proton cumulants and multi-proton correlation functions in Au+Au collisions at $\sqrt{s_\text{NN}}$=7.7, 11.5, 19.6, 27, 39, 62.4 and 200 GeV with the UrQMD model. It will provide baseline calculations of proton cumulants and correlation functions from UrQMD model to search for the QCD critical point in heavy-ion collisions.  The net-baryon fluctuations and net-proton fluctuations near QCD critical point are closely related with each other and has been discussed in many theoretical calculations such as~\cite{Kitazawa:2012at, Asakawa_formula}. At low energies, the net-proton number is dominated by protons. The discussion of conserved charge fluctuations and their correlations within the UrQMD model can be found in previous works~\cite{Xu:2016qjd, Yang:2016xga, Zhou:2017jfk}. 

The paper is organized as follow. We will first introduce the observables, the relation between the cumulants, factorial cumulants and correlation function. Then, we will discuss the rapidity, transverse momentum and energy dependence of proton cumulants and correlation function in Au+Au collisions at $\sqrt{s_\text{NN}}$=7.7 to 200 GeV from the UrQMD model. Finally, we will give a summary.

\section{Cumulants and Correlations}
In this paper, we use cumulants to characterize the fluctuations of particle multiplicities. The cumulants of event-by-event particle multiplicity distributions can be expressed in terms of moments with the following relations:
\begin{subequations}
\begin{align}
    C_1 &= \langle N \rangle \\
    C_2 &= \langle N^2 \rangle - \langle N \rangle^2 \\
    C_3 &= 2\langle N \rangle^3 - 3\langle N\rangle \langle N^2\rangle + \langle N^3 \rangle \\
    C_4 &= -6\langle N\rangle^4 + 12\langle N\rangle^2 \langle N^2\rangle - 3\langle N^2\rangle^2 \nonumber \\
        &  - 4\langle N\rangle \langle N^3\rangle + \langle N^4 \rangle
\end{align}
\end{subequations}
where the $\langle N^n \rangle$ is the $n^{th}$ order moments.

The variance, skewness and kurtosis can be calculated by cumulants, respectively:
\[
    \sigma^2    =   C_2, \quad
    S           =   \frac{C_3}{{(C_2)}^{3/2}}, \quad
    \kappa      =   \frac{C_4}{{(C_2)}^2}
\]
The ratios between different orders of cumulants can be constructed to cancel the volume when study the properties of medium. 
The $S\sigma$ and $\kappa\sigma^{2}$ can be denoted as:
\[
    S\sigma = \frac{C_3}{C_2}, \quad \kappa\sigma^2 = \frac{C_4}{C_2}
\]

On the other hand,  the cumulants can be decomposed into linear combinations of correlation functions of different order~\cite{bzdak2016cumulants,ling2016acceptance}. The correlation functions are also known as factorial cumulants~\cite{Kitazawa:2017ljq}, which are represented by the symbol $c$ in this work. The relations between cumulants and correlation functions can be written as:
\begin{eqnarray}\label{eq:c2k}
    C_1 &=& \langle N \rangle =c_1 \nonumber \\
    C_2 & = & \langle N \rangle + c_2 \nonumber \\
    C_3 & = & \langle N \rangle + 3c_2 + c_3 \nonumber \\
    C_4 & = & \langle N \rangle + 7c_2 + 6c_3 + c_4
\end{eqnarray}
and vice verse,
\begin{eqnarray} \label{eq:k2c}
    c_1 &=& \langle N \rangle =C_1 \nonumber \\
    c_2 & = & -\langle N\rangle + C_2 \nonumber \\
    c_3 & = & 2\langle N\rangle - 3C_2 + C_3 \nonumber \\
    c_4 & = & -6\langle N\rangle  + 11C_2 - 6C_3 + C_4 
\end{eqnarray}
The correlation functions can be used to probe non-Poisson distributions, since they are equal to zero for Poisson distributions.
The above relations between cumulants and correlation functions are only valid for the statistics of single variable. At lower energy where a strong enhancement of the fourth order net-proton and proton cumulant ratio ($\kappa\sigma^{2}$) is observed by the STAR experiments, the production of anti-proton is negligible. Thus, we will focus on proton cumulants and their correlation functions in Au+Au collisions from UrQMD model. Those calculations will provide insights into the signal and background contributions from various physics effects to the proton cumulants and correlation functions~\cite{Kitazawa:2017ljq,bzdak2017cumulants}, which are important to understand the experimental measurements for searching for the QCD critical point in heavy-ion collisions.

\begin{figure*}\label{fig:pdndy}
\hspace{-0.8cm}
\includegraphics[scale=0.92]{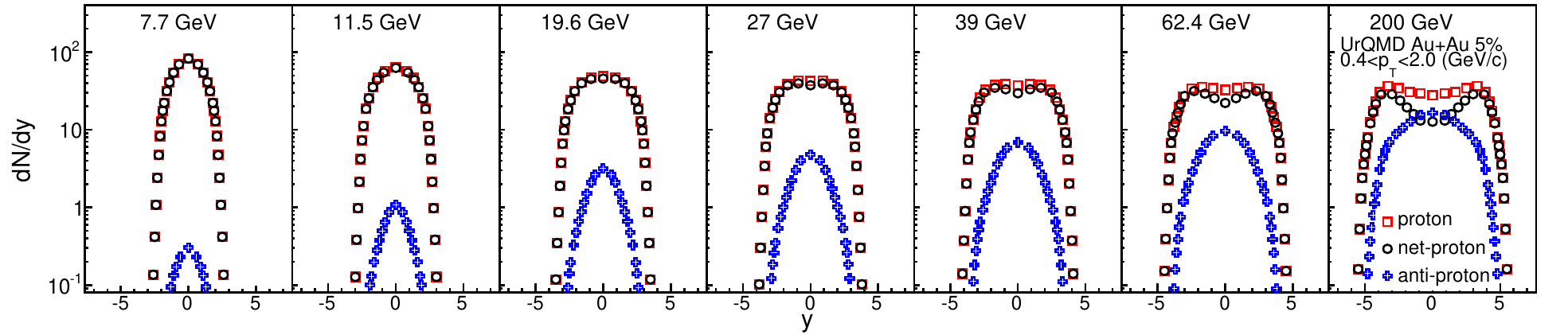}
\caption{Rapidity distributions ($\mathrm{d}N/\mathrm{d}y$) of net-proton, proton, and anti-proton in 0--5\% most central Au+Au collisions at $\sqrt{s_\text{NN}}$=7.7 to 200 GeV from UrQMD model.}
\end{figure*}

\section{\texorpdfstring{U\lowercase{r}QMD Model}{UrQMD Model}}
The UrQMD (Ultra Relativistic Quantum Molecular Dynamics) model~\cite{Bleicher:1999xi} is a well-designed transport model for the simulation with the entire available range of energies from SIS energy ($\sqrt{s_\text{NN}} = 2$ GeV) to RHIC energy ($\sqrt{s_\text{NN}} = 200$ GeV) and the collision term in the UrQMD model covers more than fifty baryon species and 45 meson species as well as their anti-particles. It is a microscopic transport model to describe hadron-hadron interactions and system evolution. Based on covariant propagation of all hadrons with stochastic binary scattering, color string formation and resonance decay, the UrQMD model provides a phase space description~\cite{bass1998microscopic} of different reaction mechanisms. At higher energies, $\sqrt{s_\text{NN}}> 5$ GeV, the quark and gluon degrees of freedom cannot be neglected. The excitation of color strings and their subsequent fragmentation into hadrons are the dominating mechanisms for the production of multiple particles. In the UrQMD, the interaction of produced particles, which may influence the acceptance of certain windows~\cite{jeon2000charged} is included. The decay of resonance may also modify the correlations~\cite{Kitazawa:2012at, bzdak2016cumulants}.

\begin{figure*}\label{fig:pcum2y}
\hspace{-0.8cm}
\includegraphics[scale=0.8]{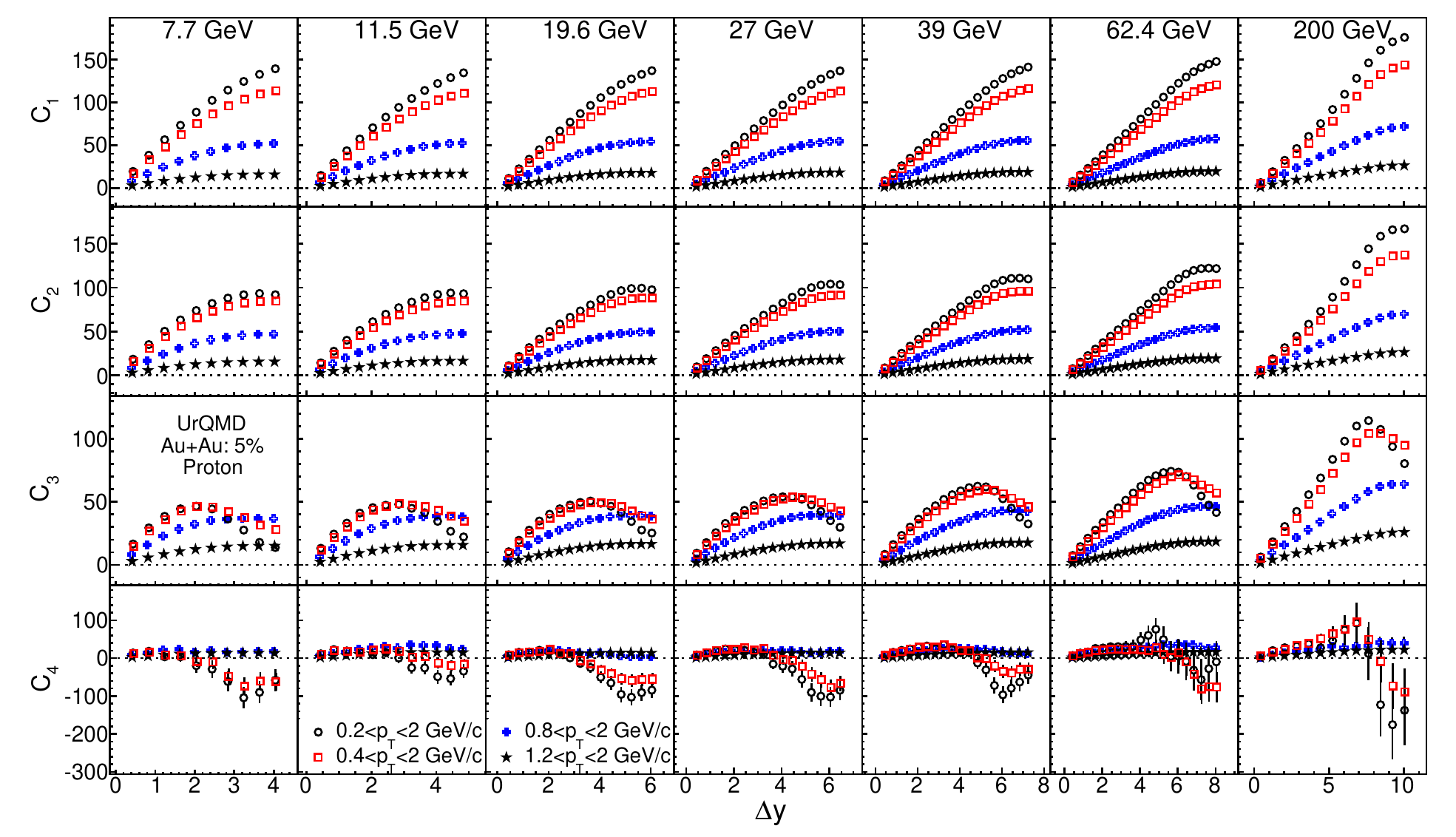}
\caption{ Acceptance (rapidity and $p_\text{T}$) dependence of the various order proton cumulants ($C_1$ to $C_4$) in 0--5\% most central Au+Au collisions at $\sqrt{s_\text{NN}}$=7.7 to 200 GeV from UrQMD model.}
\end{figure*}

\section{Results}
We analyzed the proton cumulants and correlation functions in Au+Au collisions at $\sqrt{s_\text{NN}}$=7.7, 11.5, 19.6, 27, 39, 62.4 and 200 GeV in the UrQMD model.  From low to high energies, the corresponding statistics are 35, 113, 113, 83, 135, 135 and 56 million minimum bias events, respectively.  In the model calculations, we have applied various data analysis techniques~\cite{luo2011probing, luo2013volume} to suppress the volume fluctuations and auto-correlations. Those analysis techniques have been also used in the real data analysis. For instance, to avoid auto-correlations, the collision centralities are defined by the charged pion and kaon multiplicity within pseudo-rapidity $|\eta|<1.0$. In addition, we also used the so-called Delta theorem methods~\cite{luo2012error, luo2015unified} to estimate the statistics error of cumulants and correlation functions. In our paper, we focus on the results from the 0-5\% most central Au+Au collisions events. For each energy, we studied the rapidity dependence of proton cumulants and correlation functions in different transverse momentum ($p_\text{T}$) ranges.

\begin{figure*}\label{fig:pcum2n}
\hspace{-0.8cm}
\includegraphics[scale=0.8]{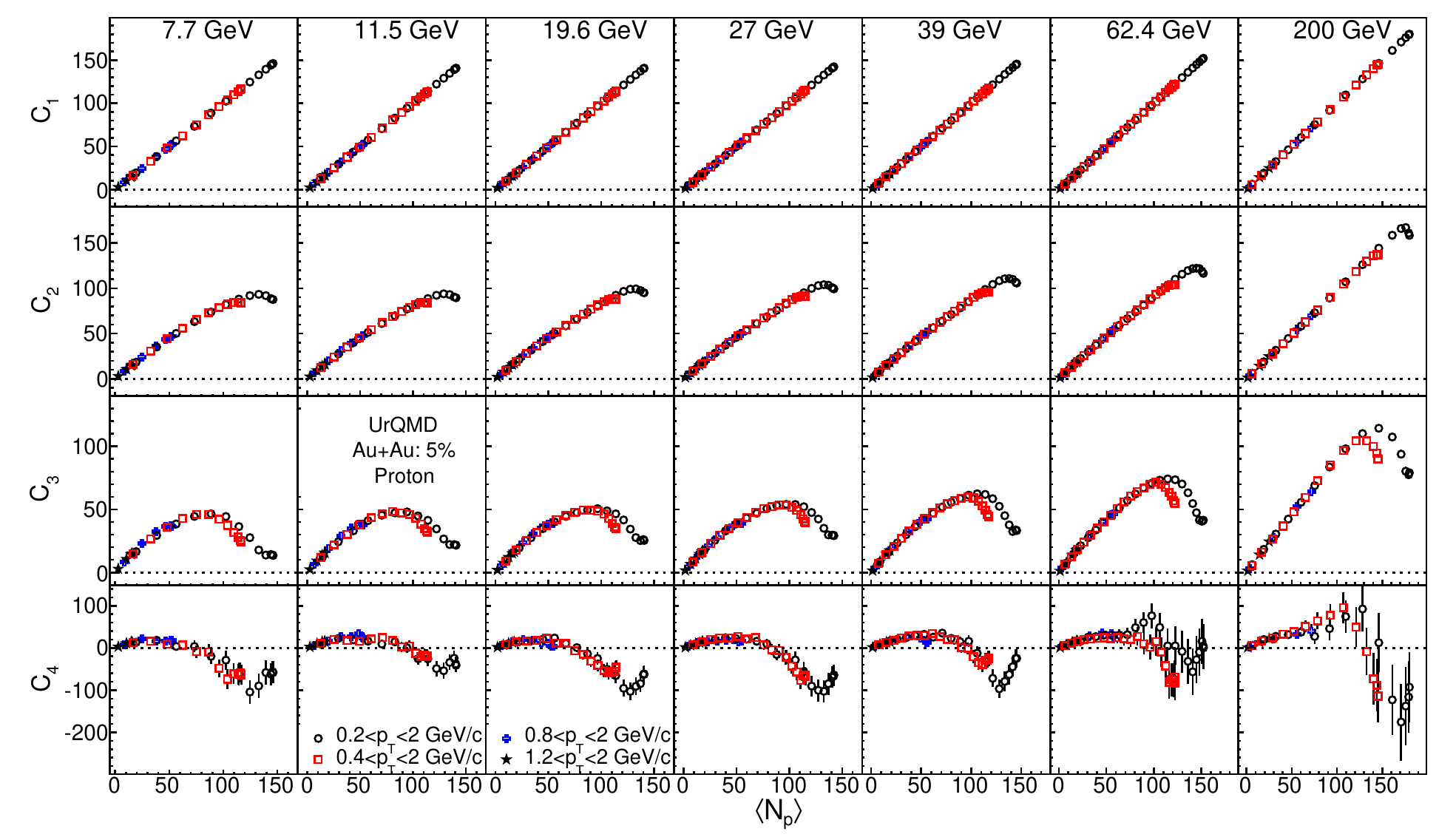}
\caption{The same as Fig.\ref{fig:pcum2y} only replace the X-axis with the corresponding mean proton number ($\langle N_p \rangle$).}
\end{figure*}

\begin{figure*}[htpb]\label{fig:pcor2y}
\hspace{-1cm}
\includegraphics[scale=0.8]{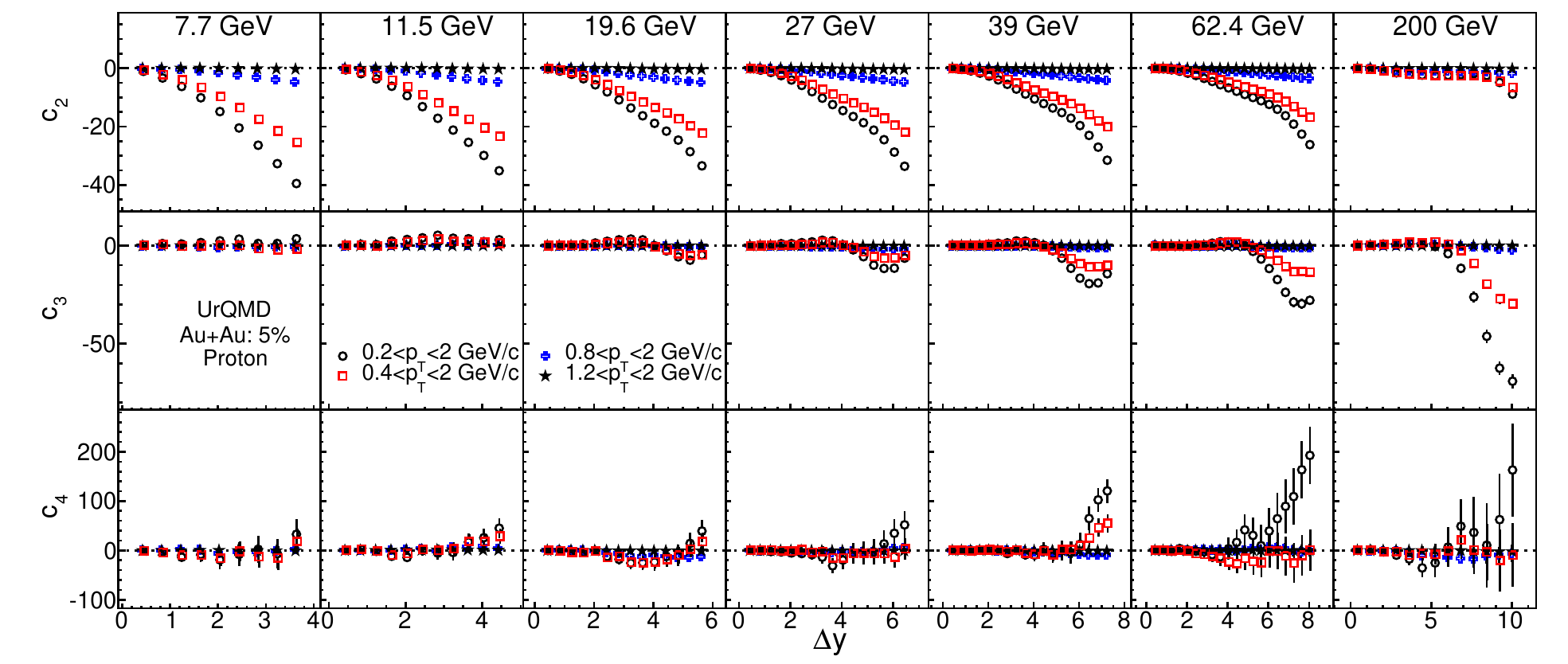}
\caption{Acceptance (rapidity and $p_\text{T}$) dependence of the proton correlation functions ($c_2$ to $c_4$) in 0--5\% most central Au+Au collisions at $\sqrt{s_\text{NN}}$=7.7 to 200 GeV from UrQMD model.}
\end{figure*}

\begin{figure*}[htpb]\label{fig:pcor2n}
\hspace{-1cm}
\includegraphics[scale=0.8]{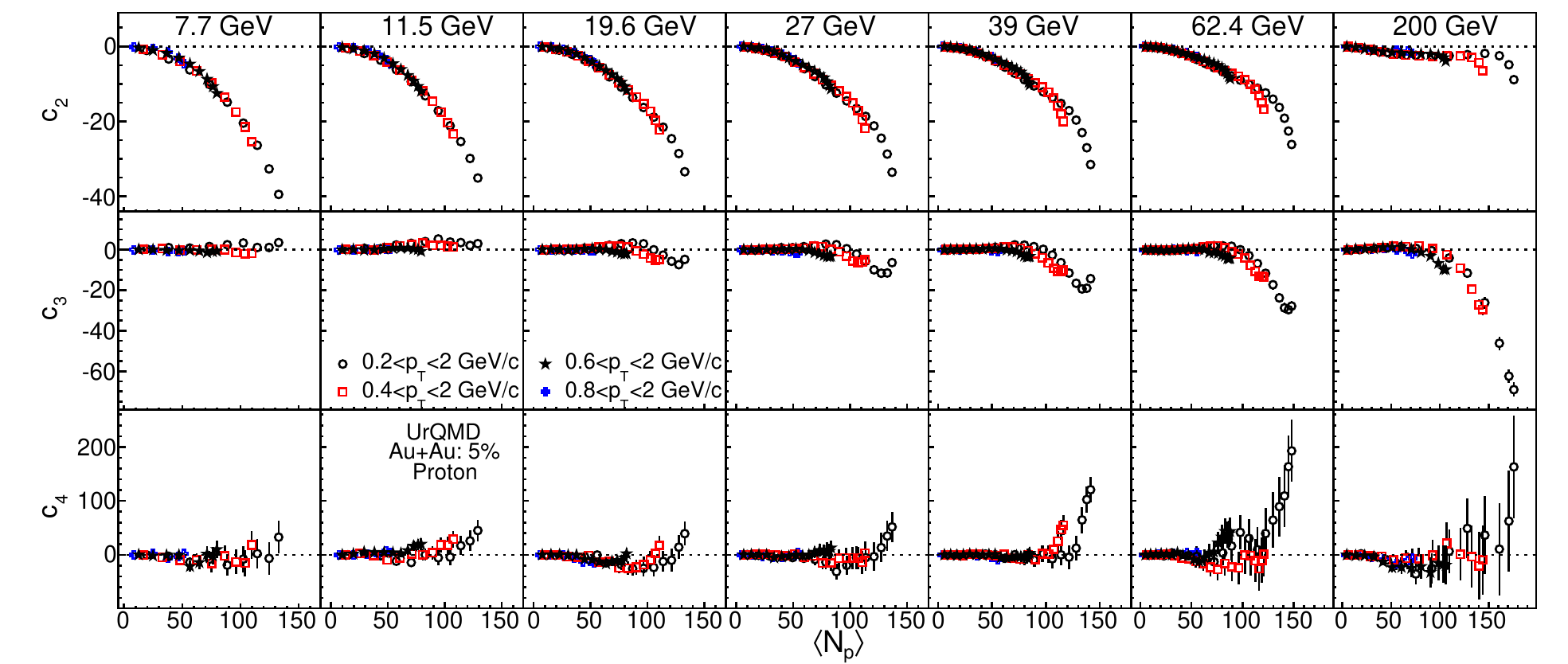}
\caption{The same as Fig.\ref{fig:pcor2y} and just replace the X-axis with the corresponding mean proton number ($\langle N_p \rangle$).}
\end{figure*}

\begin{figure*}[htpb]\label{fig:cumcor2e}
\includegraphics[scale=0.7]{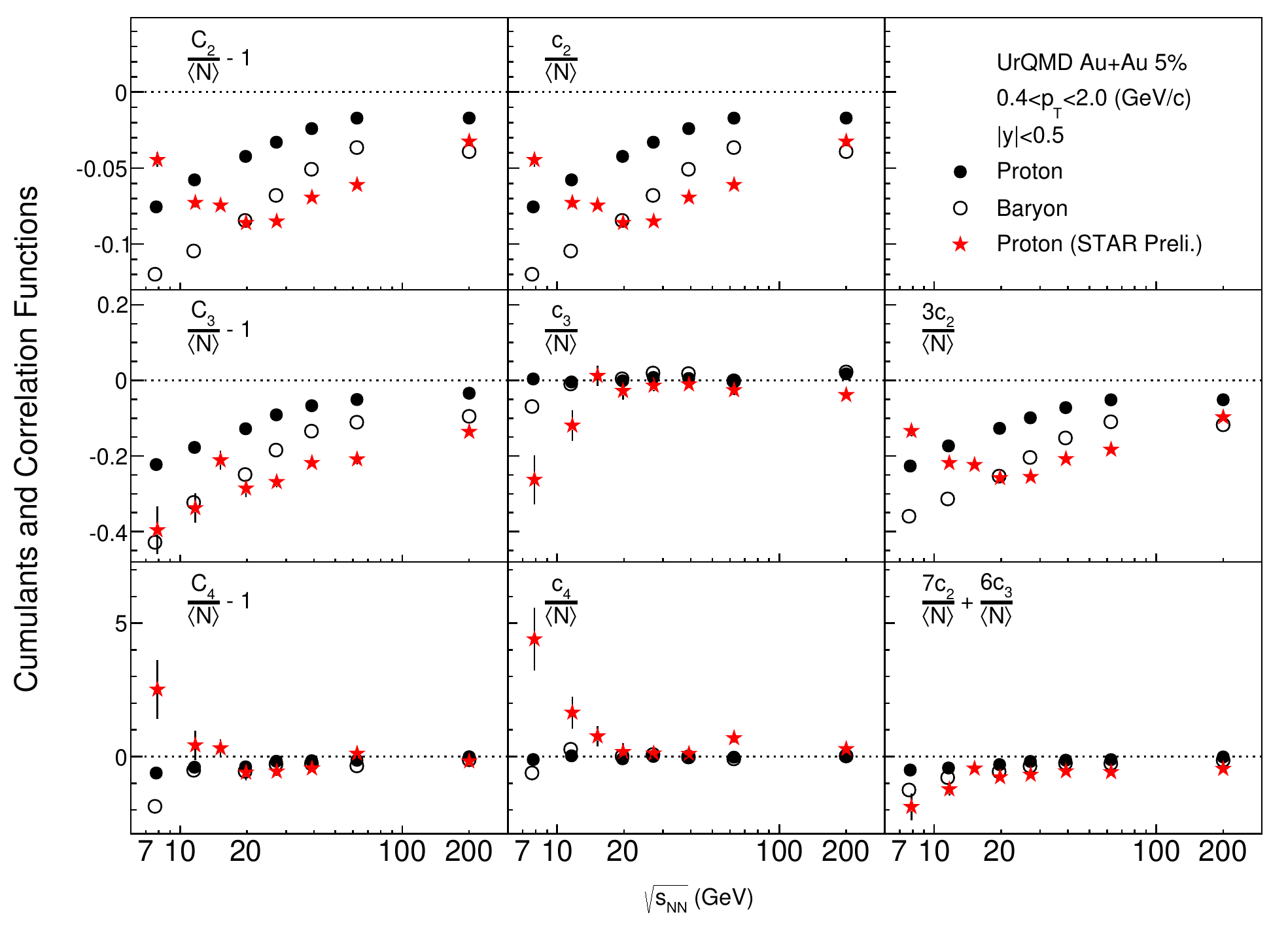}
\caption{Energy dependence of proton (baryon) cumulants and correlation functions in 0-5\% most central Au+Au collisions at $\sqrt{s_\text{NN}}$=7.7 to 200 GeV from UrQMD model (black circles) and STAR preliminary data (red stars)~\cite{Luo:2015ewa,Luo:2017faz,INT_talk}.}
\end{figure*}

Figure~\ref{fig:pdndy} shows the rapidity distribution ($\mathrm{d}N/\mathrm{d}y$) of proton, anti-proton and net-proton in the most central (0--5\%) Au+Au collisions at $\sqrt{s_\text{NN}}$=7.7 to 200 GeV in the UrQMD model. It shows that the magnitude of the proton $\mathrm{d}N/\mathrm{d}y$ distributions at $y=0$ increase monotonically as the collision energy decreases, while for anti-protons, the $\mathrm{d}N/\mathrm{d}y$ at $y=0$ increase gradually with energy increase. These can be understood by the interplay between baryon stopping and pair production of proton and anti-protons. Baryon stopping becomes more important at low energies, while pair production dominates at high energies. It is also found that the rapidity distribution of anti-proton is narrower than that of protons at each energy. The net-proton distributions follow closely with the proton distributions due to the negligible production of anti-proton at low energies, 

Figure~\ref{fig:pcum2y} shows the rapidity and $p_\text{T}$ acceptance dependence of proton cumulants in 0-5\% most central Au+Au collisions at $\sqrt{s_\text{NN}}$=7.7 to 200 GeV from UrQMD model. The $x$-axis is the rapidity window $\Delta y$, which corresponds to the rapidity coverage $(-y, y)$, and thus $\Delta y = 2y$. We plot the rapidity window within the limit of $\Delta y < 2y_\text{beam}$, where $y_\text{beam}$ is the beam rapidity of gold nuclei. From 7.7 to 200 GeV, the corresponding beam rapidities are 2.09, 2.50, 3.04, 3.36, 3.73, 4.20 and 5.36, respectively. It is found that the proton cumulants $C_{1}$ and $C_{2}$ increase monotonically with rapidity and $p_\text{T}$ acceptance, while the $C_{3}$ and $C_{4}$ show strong suppression at large acceptance. The suppressions are mainly due to the effects of baryon number conservations (BNC)~\cite{bzdak2013baryon,sakaida2014effects,Schuster:2009jv,He:2016uei}. The effects of BNC will be stronger if we have larger fraction of baryons in the acceptance. Indeed, we observed a stronger suppression for wider rapidity and/or $p_\text{T}$ acceptance. At the same acceptance coverage, the net-baryon fluctuations show larger suppression than that of net-proton (\autoref{fig:cumcor2e}). 

Figure~\ref{fig:pcum2n} shows the same results as in Fig\@.~\ref{fig:pcum2y}, we only replace the $\Delta y$ in $x$-axis with the mean proton number ($\langle N_\text{p} \rangle$) measured in the corresponding $\Delta y$ and $p_\text{T}$ cut. 
It means even if we choose the different acceptance windows, those windows which yield the same mean proton number are plotted with the same $x$ values. Interestingly, we find that the proton cumulants from different acceptance cuts (rapidity and $p_\text{T}$) show similar trends and scale with the $\langle N_\text{p} \rangle$ from various acceptance cuts within a broad energy range. 

\begin{figure*}\label{fig:prt2y}
\hspace{-0.8cm}
\includegraphics[scale=0.8]{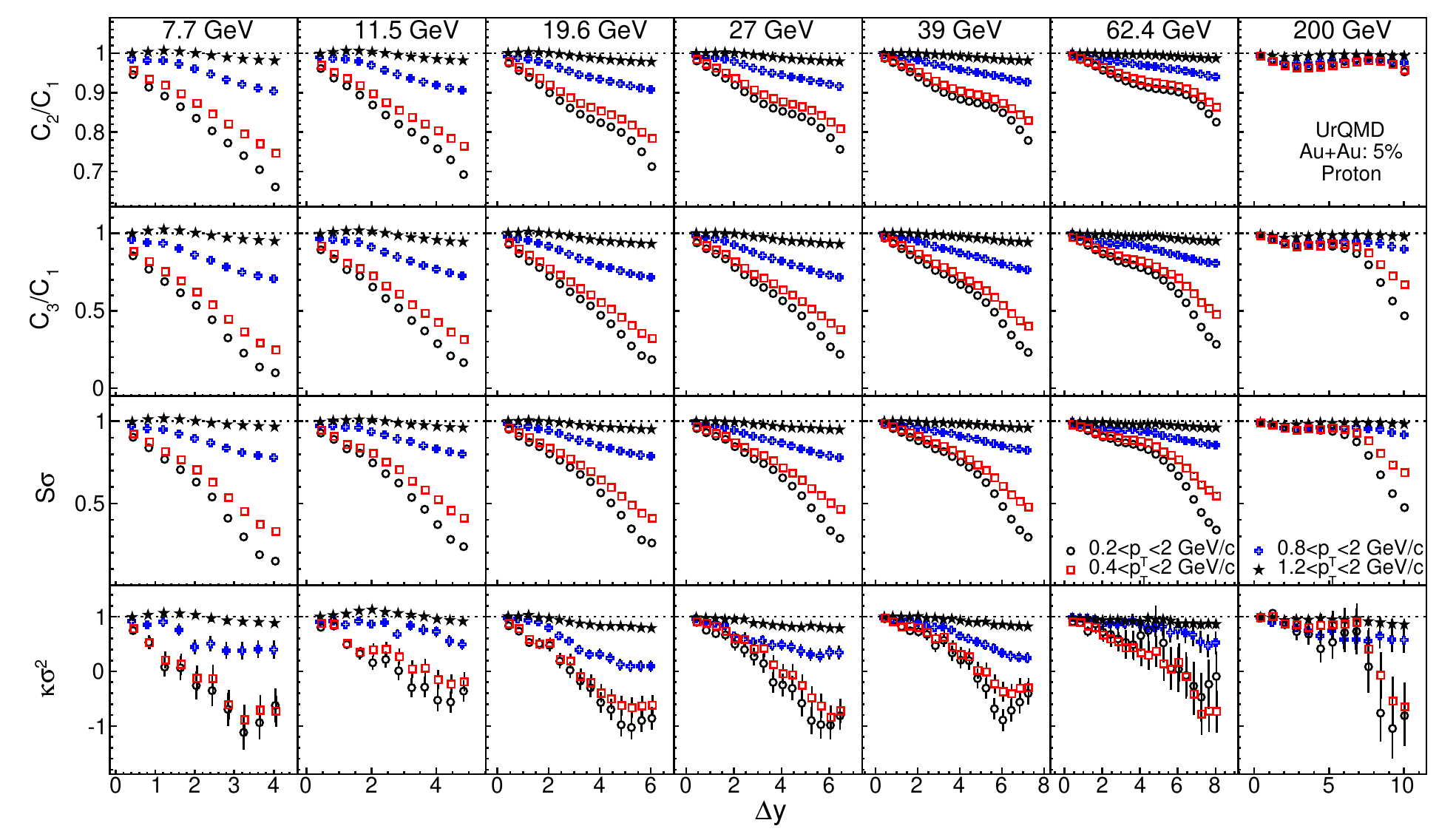}
\caption{Acceptance (rapidity and $p_\text{T}$) dependence of the various order proton cumulants ratios in 0-5\% most central Au+Au collisions at $\sqrt{s_\text{NN}}$=7.7 to 200 GeV in the UrQMD model. }
\end{figure*}

Figure~\ref{fig:pcor2y} shows the acceptance dependence for the proton correlation functions ($c_2$, $c_3$ and $c_4$) in 0-5\% most central Au+Au collisions at $\sqrt{s_\text{NN}}$=7.7 to 200 GeV in the UrQMD model. The proton correlation functions can be calculated from the proton cumulants via Eq.~\eqref{eq:k2c}. We find that the two-proton correlation ($c_2$) functions is negative and decrease monotonically when enlarging the rapidity and $p_\text{T}$ acceptance at all energies. This can be attributed to the anti-correlations between protons due to baryon number conservations~\cite{bzdak2016cumulants}. For the three ($c_3$) and four-proton ($c_4$) correlation functions, the results are close to zero and with negligible acceptance dependence. It indicates the three or four-particle correlation functions are not sensitive to the BNC. 
\begin{figure*}\label{fig:prt2n}
\hspace{-0.8cm}
\includegraphics[scale=0.8]{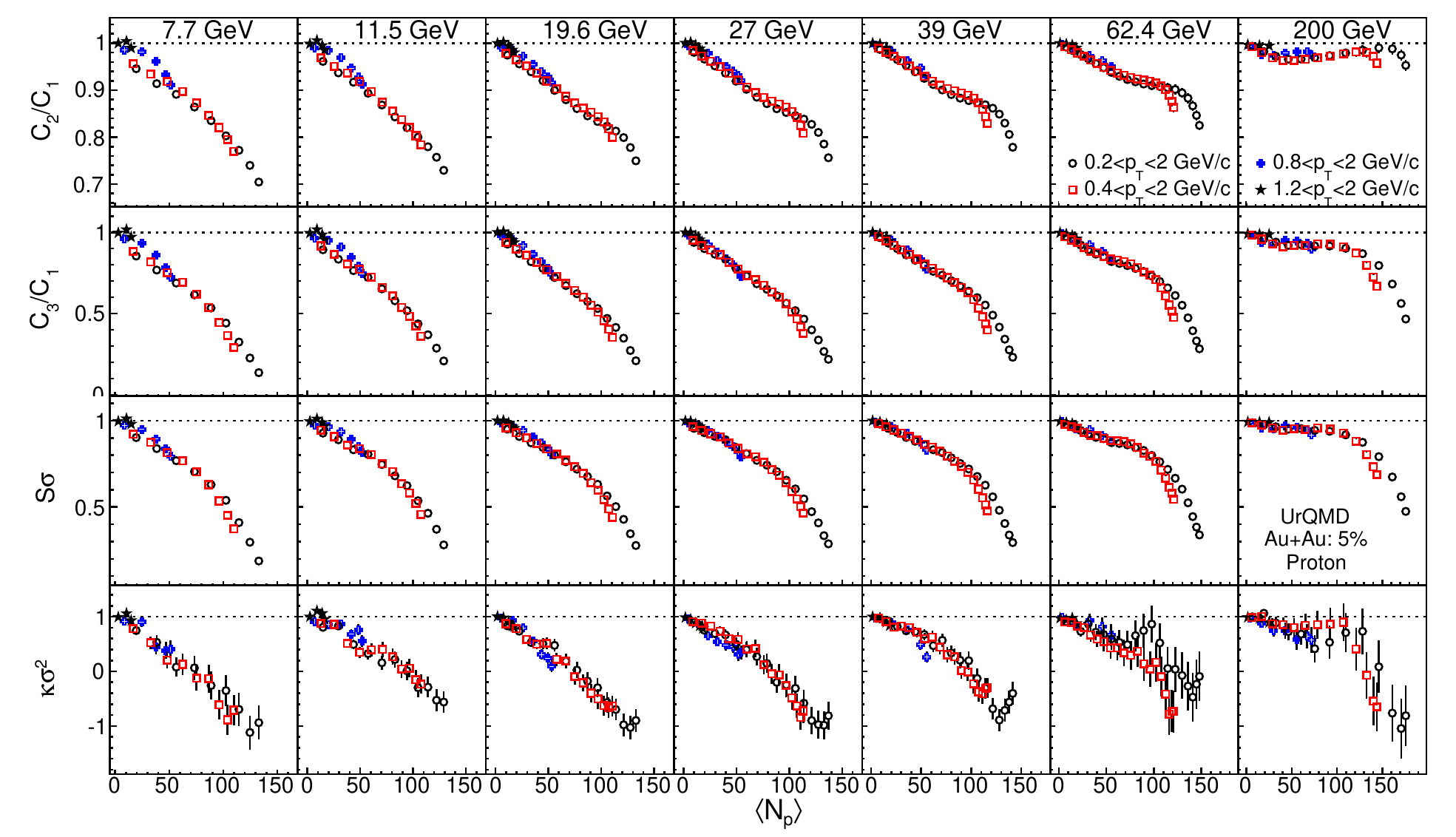}
\caption{The same as Fig.\ref{fig:prt2y} and just replace the X-axis with the corresponding mean proton number ($\langle N_p \rangle$).}
\end{figure*}

In Fig.~\ref{fig:pcor2n}, we also plot the proton correlation functions as a function of mean proton number ($\langle N_\text{p}  \rangle$), which is calculated from the corresponding acceptance (rapidity and  $p_\text{T}$). One can find that the various order proton correlation functions show scaling behavior when plotting the results as a function of the $\langle N_\text{p} \rangle$. This suggests that the proton correlation functions in Au+Au collisions in the UrQMD model are essentially scaled with mean number of protons in the acceptance, which is similar as the cumulants discussed in Fig.~\ref{fig:pcum2n}. 

To understand the scaling behavior observed in Fig~\ref{fig:pcum2n} and \ref{fig:pcor2n}, we define the reduced correlation function as:
\begin{equation}\label{eq:red}
\hat{c}_k=c_k/\langle N \rangle ^k 
\end{equation}
This parameter can be used to characterize the strength of the particle correlations in the systems. 
Based on Eq.~\eqref{eq:c2k} and~\eqref{eq:red}, we have: 
\begin{subequations}\label{eq:reds}
\begin{align}
C_2 &= \langle N\rangle +  \langle N\rangle^2\hat{c}_2 \label{eq:red2} \\ 
C_3 &= \langle N\rangle + 3\langle N\rangle^2\hat{c}_2 +  \langle N\rangle^3\hat{c}_3 \\
C_4 &= \langle N\rangle + 7\langle N\rangle^2\hat{c}_2 + 6\langle N\rangle^3\hat{c}_3 + \langle N\rangle^4\hat{c}_4 \label{eq:red22}
\end{align}
\end{subequations}
If the reduced correlation functions $\hat{c}_k$ are constant or depends only on the mean number of particles in the acceptance ($\langle N\rangle$), the cumulants and correlation functions should scale with the $\langle N\rangle$ in the various acceptance cuts. Thus, by introducing reduced correlation functions, we can easily explain the scaling behavior of the proton cumulants and correlation functions as a function of the mean number of protons. 
In the case where the particles are emitted from many uncorrelated particles sources, the reduced correlation functions are scaled as~\cite{bzdak2016cumulants}:
\begin{equation}
\hat{c}_k \propto 1/\langle N \rangle^{k-1}
\end{equation}
It means the correlation strength will be diluted by the number of particles due to the uncorrelated feature of the particle emission sources. In Poisson limit, the reduced correlation function ${{\hat c}_k} = 0$ ($k > 1$). For long-range correlations in the system, the reduced correlation functions $\hat{c_k}$ are expected to be large and constant as a function of centrality and/or acceptance. Therefore, one of the important features of the correlation function ($c_k= \langle N\rangle^k {\hat c}_k$) near the QCD critical point is the power law dependence on rapidity acceptance in heavy-ion collisions. Since the reduced correlation function doesn't depend on the binomial detector efficiency effect,  we propose to measure the rapidity dependence of the reduced proton correlation functions to probe the long-range correlations near the QCD critical point in heavy-ion collisions. 

Figure~\ref{fig:prt2y} shows the acceptance dependence of the proton cumulants ratios in Au+Au collisions at $\sqrt{s_\text{NN}}$=7.7 to 200 GeV from UrQMD model. Those ratios are constructed to eliminate the trivial volume dependence of cumulants. The expected value of the proton cumulant ratios of a Poisson distributions are unity. In our calculations, the deviations from the Poisson expectations increase with the rapidity and/or $p_\text{T}$ windows. In other words, those cumulants ratios are pushed to the Poisson expectations (unity) in the limit $\langle N \rangle \rightarrow 0$. Our calculations show that the deviations below unity become larger at low energies. This is because the effect of BNC is more important at low energies due to the stronger baryon stopping. In Fig.~\ref{fig:prt2n}, we find that the proton cumulant ratios from various acceptances follow similar trends when plotting the results as a function of the mean proton numbers. It can be understood by Eq.~\eqref{eq:red2}--\eqref{eq:red22} that the acceptance dependence should be driven by the particle multiplicity $\langle N \rangle$, if the reduced correlation functions only depend on the number of particles in the acceptance.
Figure~\ref{fig:cumcor2e} shows the energy dependence of proton (baryon) cumulants and correlation functions normalized by the mean proton numbers (also minus unity) in 0--5\% most central Au+Au collisions from both the UrQMD model and the STAR preliminary data~\cite{Luo:2015ewa,Luo:2017faz,INT_talk}. The protons in the UrQMD calculations are selected within the same rapidity and $p_\text{T}$ acceptance as used in the real data,  which are $|y| < 0.5$ and $0.4 < p_\text{T} < 2.0$ GeV/c. In the first column of the plot, we show the energy dependence of various order proton (baryon) normalized cumulants from the UrQMD and experimental data. It is found that the proton (baryon) normalized cumulants obtained from UrQMD model monotonically decrease with energy whereas the second and fourth order proton cumulants from STAR preliminary data show non-monotonic energy dependence trends, with a minimum around 20 GeV. In the UrQMD calculations, the baryon normalized cumulants show stronger suppression than that of proton. The suppression of the proton (baryon) cumulants in the UrQMD model are mainly due to the effects of BNC as explained before. The non-monotonic energy dependence of the second and fourth order proton cumulants observed in the STAR data cannot be explained by the UrQMD model without implementing critical physics. However, a caveat is that the deviation between the UrQMD and experiment might be due to non-critical correlations that are not captured by the UrQMD\@.

To understand the contributions to the cumulants from different physics effects, we decompose the various order cumulants into multi-particle correlation functions based on the equations~\eqref{eq:c2k} and~\eqref{eq:k2c}. It means that each cumulant in the first column is just equal to the sum of the results in the second and the third columns. It is easily noticed that the strong suppression observed in various order proton (baryon) cumulants from UrQMD at low energies are mainly caused by the negative two-proton correlation functions ($c_2$), which is due to the anti-correlation between proton (baryon) caused by the BNC effects. The results for the three and four-particle correlation functions for protons (baryons) in the UrQMD model show a flat energy dependence and close to zero. It indicates that the high order ($>2$) proton (baryon) correlation functions are not sensitive to the effect of BNC, which could serve as a better probe of the critical fluctuations in heavy-ion collisions. 

On the experimental side, the contributions of two-proton correlation functions cannot explain the suppression of the third order and large increase of the fourth order proton cumulants at low energies measured by the STAR experiment. The suppression and enhancement of the third and the fourth order proton cumulants are due to the suppression and enhancement of the three and four-proton correlation functions with respect to zero, respectively. Most importantly, it is found that the STAR preliminary results of non-monotonic energy dependence of the fourth order proton cumulants are dominated by the four-proton correlation functions, which requires cluster formation caused by multi-proton stopping or a phase transition~\cite{bzdak2016cumulants,bzdak2017cumulants}.

\section{Summary}
We conducted a numerical calculation with the UrQMD model to study the acceptance dependence of the proton cumulants and correlation functions in 0-5\% most central Au + Au collisions at $\sqrt{s_\text{NN}}$=7.7-200 GeV. We observed strong suppressions in the acceptance dependence of high order proton cumulants. Those suppressions can be understood in terms of the negative two-proton correlation functions caused by the baryon number conservations. However, the values of three and four-proton correlation functions are close to zero and have negligible acceptance dependence. It suggests that the three and four-proton correlation functions could serve as better probes of the critical fluctuations, since they are sensitive to correlation lengths, but less affected by the baryon number conservations. By comparing the UrQMD results with experimental data, we found that the second and fourth order proton cumulants measured by STAR experiment show a non-monotonic energy dependence, with a minimum around 20 GeV and a significant enhancement at lower energies. However, the results from the UrQMD model monotonically decrease with energy and cannot explain the trends observed in STAR data. The non-monotonic energy dependence observed in the fourth order proton cumulants are dominated by the four-proton correlation functions. 

In the UrQMD calculations, the proton cumulants and correlation functions have similar trends and show scaling behavior as a function of mean number of protons. This indicates that the reduced proton correlation functions $\hat{c}_k$ could be constant or only depend on the mean number of protons in the acceptance. On the other hand, the reduced proton correlation functions are sensitive to the correlation length and less affected by the baryon number conservations in heavy-ion collisions. Furthermore, the binomial detector efficiency effect will be canceled out in the reduced correlation functions. Thus, we propose to measure the rapidity dependence of the reduced proton correlation functions to probe the long-range correlation near the QCD critical point.

\section*{Acknowledgment}
The work was supported in part by the MoST of China 973-Project No.2015CB856901, NSFC under grant No. 11575069.  

\bibliography{ref}
\end{document}